\documentclass[aps,prb,twocolumn,floatfix,superscriptaddress]{revtex4-2}%
\usepackage{graphicx}
\usepackage{amssymb}
\usepackage{graphicx}
\usepackage{natbib}
\usepackage{amsmath}
\usepackage{amsfonts}
\usepackage{color}
\providecommand{\U}[1]{\protect\rule{.1in}{.1in}}

\begin{document}

\title{Optical read-out of the N\'eel vector in metallic antiferromagnet Mn$_{2}$Au}
\author{Vladimir Grigorev}
\affiliation{Institute of Physics, Johannes Gutenberg University, 55128 Mainz, Germany}
\affiliation{Graduate School of Excellence Materials Science in Mainz, 55128 Mainz, Germany}
\author{Mariia Filianina}
\affiliation{Institute of Physics, Johannes Gutenberg University, 55128 Mainz, Germany}
\affiliation{Graduate School of Excellence Materials Science in Mainz, 55128 Mainz, Germany}
\author{Stanislav Yu. Bodnar}
\affiliation{Institute of Physics, Johannes Gutenberg University, 55128 Mainz, Germany}
\affiliation{Walter Schottky Institut and Physics Department, Technische Universität München, 85748 Garching, Germany}
\author{Sergei Sobolev}
\affiliation{Institute of Physics, Johannes Gutenberg University, 55128 Mainz, Germany}
\author{Nilabha Bhattacharjee}
\affiliation{Institute of Physics, Johannes Gutenberg University, 55128 Mainz, Germany}
\author{Satya Bommanaboyena}
\affiliation{Institute of Physics, Johannes Gutenberg University, 55128 Mainz, Germany}
\author{Yaryna Lytvynenko}
\affiliation{Institute of Physics, Johannes Gutenberg University, 55128 Mainz, Germany}
\author{Yurii Skourski}
\affiliation{Dresden High Magnetic Field Laboratory (HLD-EMFL), Helmholtz-Zentrum
Dresden-Rossendorf, 01328 Dresden, Germany}
\author{Dirk Fuchs}
\affiliation{Institute for Quantum Materials and Technologies (IQMT), Karlsruhe Institute
of Technology, 76344 Eggenstein-Leopoldshafen, Germany}
\author{Mathias Kl\"aui}
\affiliation{Institute of Physics, Johannes Gutenberg University, 55128 Mainz, Germany}
\affiliation{Graduate School of Excellence Materials Science in Mainz, 55128 Mainz, Germany}
\author{Martin Jourdan}
\affiliation{Institute of Physics, Johannes Gutenberg University, 55128 Mainz, Germany}
\author{Jure Demsar}
\affiliation{Institute of Physics, Johannes Gutenberg University, 55128 Mainz, Germany}
\affiliation{Graduate School of Excellence Materials Science in Mainz, 55128 Mainz, Germany}

\date{\today}

\begin{abstract}
Metallic antiferromagnets with broken inversion symmetry on the two
sublattices, strong spin-orbit coupling and high N\'{e}el temperatures offer new opportunities for applications in spintronics. Especially Mn$_{2}$Au, with high N\'{e}el temperature and conductivity, is particularly interesting for real-world applications. Here, manipulation of the orientation of the staggered magnetization,\textit{\ i.e.} the N\'{e}el vector, by current pulses has been recently demonstrated, with the read-out limited to studies of anisotropic magnetoresistance or X-ray magnetic linear dichroism. Here, we report on the in-plane reflectivity anisotropy of Mn$_{2}$Au (001) films, which were N\'{e}el vector aligned in pulsed magnetic fields. In the near-infrared, the anisotropy is $\approx$ 0.6\%, with higher reflectivity for the light polarized along the N\'{e}el vector. The observed magnetic linear dichroism is about four times larger than the anisotropic magnetoresistance. This suggests the dichroism in Mn$_{2}$Au is a result of the strong spin-orbit interactions giving rise to anisotropy of interband optical transitions, in-line with recent studies of electronic band-structure. The considerable magnetic linear dichroism in the near-infrared could be used for ultrafast optical read-out of the N\'{e}el vector in Mn$_{2}$Au.
\end{abstract}

\maketitle

\section{Introduction}

There is an increasing interest in antiferromagnetic (AFM) materials as active
elements in future spintronics devices, including data storage applications
\cite{spintronics_1,spintronics_2,spintronics_3,spintronics_4,spintronics_5}.
This is motivated by the absence of net magnetization and the related stray
fields, which limit the minimum distance between two bits, and thus leads to
higher storage density compared to ferromagnets. Moreover, the inherently
fast spin dynamics in AFMs, which takes place in the THz range
\cite{Kampfrath_THz,Reid_APL_THz,Yamaguchi_PRL_THz,Kim_2014_THz, Arena_THz},
is orders of magnitude faster then typical spin dynamics in ferromagnetic
materials (GHz range) \cite{GHz_dynamics}. The absence of net magnetization,
however, makes the manipulation and read-out of magnetic order, which is given
by the staggered magnetization, \textit{i.e.} the N\'{e}el vector, generally difficult.

Recently, two fully compensated metallic antiferromagnets CuMnAs and Mn$_{2}%
$Au have been in the research focus. Here, the specific crystal and magnetic
structure in combination with large spin-orbit coupling enable current driven
manipulation of the N\'{e}el vector \cite{Zelezny}. In particular, it was
suggested that application of electric current pulses of the order of
$10^{8}-10^{9}$A cm$^{-2}$ can rotate the N\'{e}el vector due to the bulk
N\'{e}el spin-orbit torques (NSOT) \cite{Zelezny}. Such current-driven
switching was indeed demonstrated in metallic AFMs CuMnAs
\cite{Wadley_Sci_2016,CuMnAs_PRL,Wadley_imaging} and Mn$_{2}$Au
\cite{Stas_Nature,Bodnar_PRB,electrical_switch_thermal}, with pulse durations
down to picoseconds, achieved by driving currents with pulsed THz radiation
\cite{OleTHz}. Note that the high N\'{e}el temperature ($\approx1500$ K) and
high electrical conductivity \cite{Barthem,Jourdan_growth} make Mn$_{2}$Au
particularly interesting for real-world applications. In Mn$_{2}$Au the
current pulses were shown to reorient only a fraction (up to 30 $\%$) of the
domains \cite{Bodnar_PRB}, implying further studies of the switching
mechanisms as well as novel approaches are required.

The absence of net magnetization makes the efficient read-out of the direction
of the N\'{e}el vector challenging. Indeed, the electrical read-out via the
anisotropic magnetoresistance (AMR) and the planar Hall effect is well
established \cite{Wadley_Sci_2016,Stas_Nature}. However, recent studies show
that in Mn$_{2}$Au the AMR does not exceed 0.15 \% \cite{Bodnar_AMR}.
Moreover, the read-out time in the AMR detection scheme is inherently limited
by the electronics to timescales orders of magnitude larger than the intrinsic
switching timescales \cite{OleTHz}. Thus, using optical methods, which would
enable fast read-out on the femtosecond scale, should be explored
\cite{Nemec,Siddiqui}. While linear magneto-optic effects are commonly used to
investigate magnetization and its dynamics in ferro- and ferri-magnets
\cite{Erskine,Kimel}, as well as in non-collinear antiferromagnets \cite{Higo},
the presence of the quadratic magneto-optical effects (Cotton-Mouton/Voigt
effect) was demonstrated in several insulating AFMs
\cite{Borovick-Romanov,Pisarev_MLD,Ferre_MLD,Reid_APL_THz}. To probe
the small changes in the refractive index due to the quadratic magneto-optical
effects, polarimetry studies are commonly performed in the transmission
geometry, since the changes in the optical phase accumulate over the optical
path length within the material \cite{Borovick-Romanov,Pisarev_MLD,Ferre_MLD}.
For metallic collinear AFM, with optical penetration depths on the order of
tens of nanometers, the corresponding changes are small. Thus, for
\nolinebreak{CuMnAs} another approach has been recently demonstrated
\cite{CuMnAs_pp,Saidl2020}. Here, the thermomodulation aspect of the optical
pump-probe technique has been used to detect small photoinduced changes in rotation
of the polarization of the optical probe beam. Since such changes depend on
the polarization of light with respect to the N\'{e}el vector, probe
polarization dependence is used to determine the direction of the N\'{e}el
vector \cite{CuMnAs_pp,Saidl2020}. Indeed, similar approach has been recently
used to study dynamics in metallic AFM Fe$_{2}$As \cite{Yang} and insulating
CoO \cite{CoO_pp}. Finally, we note, that several recent reports demonstrated large magneto-optical contrast also in ultrathin NiO \cite{Xu19,Schreiber,Meer} and CoO \cite{Xu20} films. However, in this case the origin of the magneto-optical contrast is in large magnetostriction.

In Mn$_{2}$Au, the recent angular-resolved photoemission study of N\'{e}el
vector aligned Mn$_{2}$Au films demonstrated the breaking of the C$_{4z}$
symmetry, a consequence of antiferromagnetic order and strong spin-orbit
interaction \cite{harpes_mn2au}. In fact, a pronounced in-plane anisotropy of
the electronic band structure is observed up to the binding energy of a few eV
\cite{harpes_mn2au}. The existence of flat bands at the 1.5 eV binding energy,
and their calculated anisotropy may present means to detect a substantial
magnetic linear dichroism in the near-infrared (NIR) range.

Here, we report on near-infrared magnetic linear dichroism (MLD) measurements on Mn$_{2}$Au thin films, performed in reflectance geometry at room temperature. To determine the magnitude of the MLD we investigate Mn$_{2}$Au films whose staggered
magnetization (N\'{e}el vector) was aligned using pulsed magnetic field
\cite{Bodnar_AMR}. The MLD in the NIR range is found to be $\approx0.6$ \%,
about four times larger than the observed anisotropic magnetoresistance in
Mn$_{2}$Au \cite{Bodnar_AMR}. The comparatively large MLD suggests it
originates from the anisotropy in the interband absorption as a result of
C$_{4z}$ symmetry breaking.

\section{Results and discussion}

\subsection{Mn$_{2}$Au thin films\label{Films}}

Mn$_{2}$Au has a body-centered tetragonal crystal structure, whose unit cell
is depicted in Fig. 1(d). It is an easy plane (001) AFM, with a strong out-of
plane hard axis and a weak in-plane magnetic anisotropy. The spin orientations
in adjacent layers are presented by arrows, with the N\'{e}el vector pointing
along the easy [110] directions \cite{Barthem,Jourdan_growth,FilmsAFM}.

The $c$-axis epitaxial Mn$_{2}$Au thin film is grown on r-cut (1$\overline{1}%
$02) Al$_{2}$O$_{3}$ substrate, with the lateral size of 10$\times$10 mm$^{2}$
and thickness of 530 $\mu$m by the radio--frequency magnetron sputtering at
600 $^{o}$C - see Ref. [21,37] for fabrication details. To ensure epitaxial
growth, a 45 nm thick Mn$_{2}$Au film is deposited on a 20 nm thick (001) Ta
buffer layer. To protect the surface, a 2 nm Al layer is deposited on Mn$_{2}%
$Au, forming an aluminum-oxide capping layer. Mn$_{2}$Au grows epitaxially
with [110] and [1$\overline{1}$0] axes parallel to the substrate edges, which
are along the $[010]_{s}$ and $[211]_{s}$ directions of  r-cut Al$_{2}$O$_{3}$. As
elaborated in Appendix B, the $c$-axis of Mn$_{2}$Au films grown on
r-cut Al$_{2}$O$_{3}$ is tilted by 2-3$^{o}$ towards the $[010]_{s}$ direction.

As previously demonstrated \cite{Leha_PRB,Leha_PSS,Bodnar_AMR,harpes_mn2au},
the application of magnetic field, exceeding 30-50 T results in the aligning
the N\'{e}el vector along one of the easy axes, perpendicular to the applied
magnetic field \cite{Leha_PRB}. While such a "polarization" of the film may be
incomplete, \textit{i.e.} there is still a small fraction of orthogonally
polarized domains, it is quasi permanent
\cite{harpes_mn2au,Leha_PRB,Bodnar_AMR}.

To perform systematic measurements, the as-grown sample is cut into 5$\times$5
mm$^{2}$ pieces. The three pieces investigated are labeled with letters
$A$-$C$. Sample $A$ is in the as-prepared state, \textit{i.e.} in a
multidomain configuration, and is used as a reference. Samples $B$ and $C$ are
N\'{e}el vector aligned \cite{harpes_mn2au,Leha_PRB,Bodnar_AMR} in the pulsed
magnetic field of 60 T along the different, yet crystallographically
equivalent \{110\} directions, as shown in the middle panel of Fig. 1.
Throughout this work the default film orientation is such that $x\parallel
\lbrack010]_{s}$ and $y\parallel\lbrack211]_{s}$.

\begin{figure}[htp]
\centerline{\includegraphics[width=90mm]{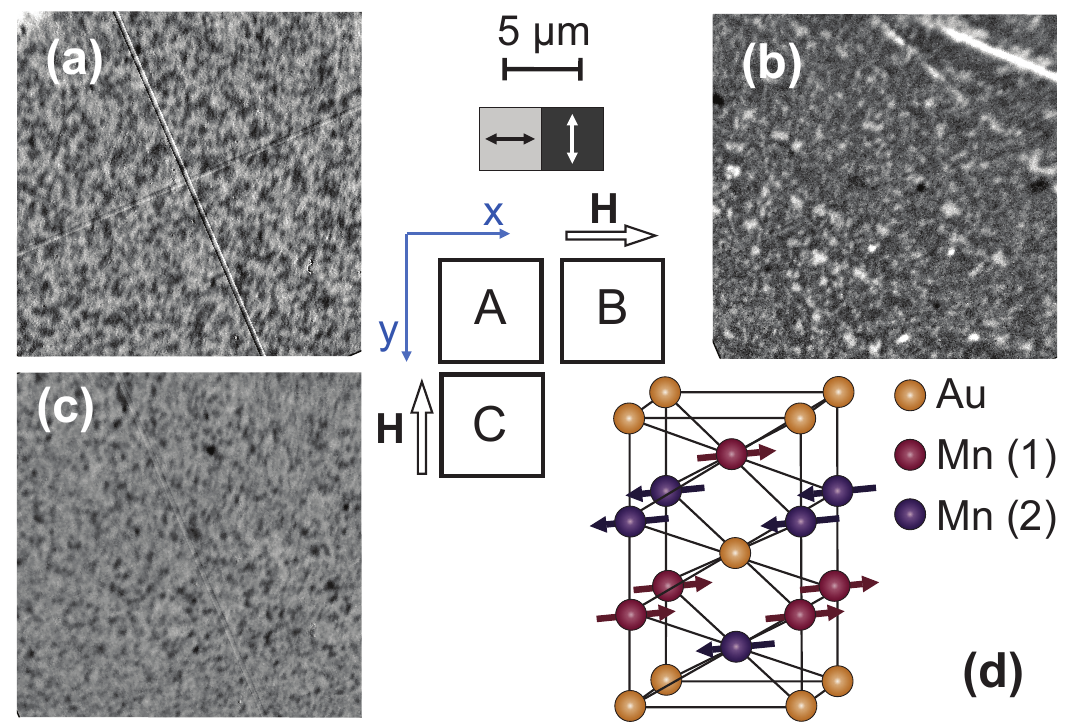}} \caption{The sample layout
and the XMLD-PEEM images of samples $A$-$C$, shown in panels (a)-(c),
respectively. Mn$_{2}$Au grows epitaxially with [110] and [1$\overline{1}$0]
axes parallel to the substrate edges which are along $[010]_{s}$ ($x$) and
$[211]_{s}$ ($y$) axes of the r-cut sapphire. Sample $A$ was as grown, while
samples $B$ and $C$ were orthogonally polarized in the 60 T pulsed magnetic field. The directions of the applied magnetic field and the corresponding
N\'{e}el vector directions are shown in the center, together with the scale of the XMLD-PEEM images. The contrast shows the two types of magnetic domains
with the N\'{e}el vector along the two orthogonal easy axes: [110] and [1$\overline{1}$0]. In the as-prepared sample $A$, roughly equal volume fraction
of both types of domains is observed, while the aligned samples $B$ and $C$ display mostly one type of the domains. (d) The unit cell of Mn$_{2}$Au, with spin orientations in adjacent layers shown by arrows.}%
\label{PEEM}%
\end{figure}

\subsection{X-ray magnetic linear dichroism in polarized Mn$_{2}$Au films}

Prior to optical studies the AFM domain patterns of all samples are first
imaged by X-ray magnetic linear dichroism photoemission electron microscopy
(XMLD-PEEM). The XMLD-PEEM imaging of the AFM domains is performed at room temperature at the
SIM beamline of the Swiss Light Source. The sample is illuminated by linearly
polarized X-rays with the polarization in the sample plane and the angle of
incidence of 16$^{\circ}$. The XMLD contrast was obtained by using the
two-energy mode, as described in detail in Ref.[39]. Here the energies E$_{1}$
= 639 eV and E$_{2}$ = 638.2 eV were used, which correspond to the maximum and
minimum of the XMLD \cite{Leha_PSS}.

The XMLD-PEEM images of all samples are shown in Fig. 1, demonstrating that
samples $B$ and $C$ have orthogonally oriented N\'{e}el vector. In particular,
each of the polarized samples has mostly one type of AFM domains, with narrow
worm-like domains of the other type, similarly to earlier reports
\cite{Leha_PRB,Leha_PSS}. The sample $A$, on the other hand, has about equal
density of the domains of both types. Hence, sample $A$ is used as a reference
in optical dichroism measurements.

\subsection{Magnetic linear dichroism in the near-infrared}

\subsubsection*{General symmetry considerations}

In a tetragonal easy-plane system like Mn$_{2}$Au a symmetry-based
phenomenological model of the quadratic magneto-optical effect generally
implies two possible contributions to the MLD: an isotropic and an anisotropic
\cite{Borovick-Romanov}. The isotropic contribution does not depend on the
direction of the N\'{e}el vector within the plane, and can be observed when
light is propagating along the two-fold symmetry axis of the crystal (here, we
are considering the crystal symmetry only). The anisotropic MLD does depend on
the direction of the N\'{e}el vector \cite{Pisarev_MLD} and is present when
the light propagates along the fourfold symmetry axis. It is the anisotropic
component, that can be used to determine the N\'{e}el vector orientation in a
collinear AFM.

The corresponding dielectric tensor of an AFM with the tetragonal crystal
structure and the N\'{e}el vector perpendicular to the fourfold symmetry axis
(\textit{i.e.} within the easy $a-b$ plane), can be written as
\cite{Borovick-Romanov}:%

\begin{align}
&  \epsilon_{xx}=\epsilon_{\bot}+\lambda_{1}\vec{L}^{2} +\frac{1}{2}%
\lambda_{3}(L_{x}^{2}-L_{y}^{2})-2\lambda_{4}L_{x}L_{y}\nonumber\\
&  \epsilon_{yy}=\epsilon_{\bot}+\lambda_{1}\vec{L}^{2} -\frac{1}{2}%
\lambda_{3}(L_{x}^{2}-L_{y}^{2})+2\lambda_{4}L_{x}L_{y}\nonumber\\
&  \epsilon_{zz}=\epsilon_{||}+ \lambda_{2}\vec{L}^{2}\\
&  \epsilon_{xy}=\epsilon_{yx}= \frac{1}{2}\lambda_{3}(L_{x}^{2}-L_{y}%
^{2})+2\lambda_{4}L_{x}L_{y}\nonumber\\
&  \epsilon_{xz}=\epsilon_{zx}=\epsilon_{yz} =\epsilon_{zy}=0
\nonumber\label{Eq1}%
\end{align}

Here $\epsilon_{\bot}$ and $\epsilon_{||}$ are the in-plane and out-of-plane
dielectric constants in the absence of magnetic order, $\vec{L}$ is the
N\'{e}el vector and $\lambda_{i}$ are the phenomenological coefficients
related to the MLD. Given the two possible directions of the N\'{e}el vector
in Mn$_{2}$Au, we choose the $z-$axis to be along the fourfold [001] axis,
while $x$ and $y$ coincide with [110] and [1$\overline{1}$0] crystallographic
axes. The difference between $\epsilon_{xx/yy}$ and $\epsilon_{zz}$ is
governed by $\epsilon_{\bot}-\epsilon_{||}$, yet includes also the component
proportional to $\vec{L}^{2}$, \textit{i.e.} the so called isotropic component
to the MLD (the term proportional to $\vec{L}^{2}$ is likely negligible
compared to $\epsilon_{\bot}-\epsilon_{||}$). The difference $\epsilon_{\bot
}-\epsilon_{||}$ may lead to the reflectivity-anisotropy in non-normal
incidence configuration (see Appendix B).

\subsubsection*{Experimental approach}

The goal of the experiment is to determine the anisotropic magnetic
contribution to dichroism, given by $\epsilon_{xx}-\epsilon_{yy}$, which is
proportional to phenomenological constants $\lambda_{3}$ and $\lambda_{4}$ and
depends on the (in-plane) N\'{e}el vector orientation. Such a MLD is expected
to be in the range of 10$^{-3}$ and could be overshadowed by extrinsic
non-magnetic effects such as strain. In particular, in the case of an
additional lattice distortion in the $ab$-plane due to strain, $\epsilon
_{\bot}$ (and $\lambda_{1}$ as well) would be different for the $x$ and $y$
directions, leading to reflectivity-anisotropy in the $ab$-plane unrelated to
the magnetic order (see the discussion in Appendix B). Furthermore, such
extrinsic contributions may vary from sample to sample due to slight variation
of substrate properties such as miscut. Thus, to unambiguously determine the
MLD, measurements on different magnetically aligned parts of the same film are performed.

\begin{figure}[htp]
\centering \includegraphics[width=0.95\columnwidth]{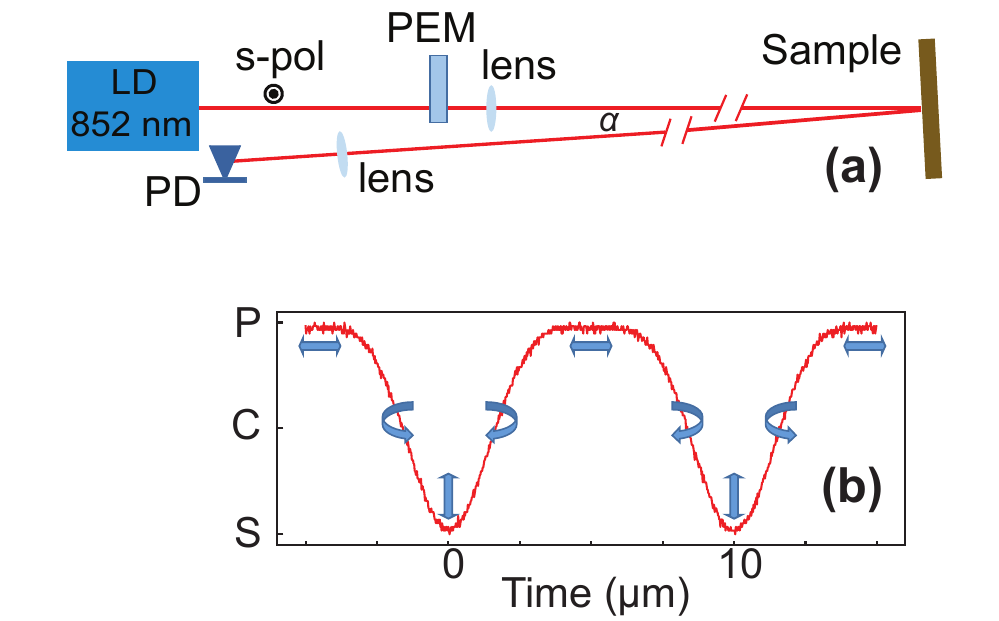}\caption{(a)
Schematic of the experimental setup (top view). The laser diode (LD) is vertically
polarized ($s$-polarized). The polarization is modulated by the photo-elastic modulator (PEM)
operating at 50 kHz. The distance between the LD and the sample is about 2.5
m, while the photodiode (PD) is mounted about 3 cm from the laser output,
giving the incidence angle $\alpha/2\approx0.34^{o}$. Panel (b)
presents the time evolution of the light intensity on the PD, when a polarizer
is placed in front of the PD selecting the horizontal ($p$-polarized) signal
(the modulation period is 10 $\mu$s). The blue arrows represent the
polarization states at given times, varying between vertical ($s$) and
horizontal ($p$) polarizations that are along the [110]/[1$\overline{1}$0]
crystalline axes of Mn$_{2}$Au.}%
\label{Setup}%
\end{figure}

Optical dichroism is studied using a setup shown in Fig. 2(a) - for details see Appendix A. The measurements are performed at room temperature. We use the 852 nm laser diode and a Photo-Elastic modulator (PEM) to modulate the polarization of incoming light between the two orthogonal linear polarizations: \textit{s-}polarized (vertical) and \textit{p-}polarized (horizontal). The Mn$_{2}$Au films were mounted such that their easy axes are along the horizontal/vertical directions, which correspond to $x$ and $y$ directions of $\epsilon_{ij}$. The light reflected from the sample is focused onto a photodiode and the AC signal is recorded using a digital oscilloscope. Measurements at multiple positions on each of the three samples are performed (the laser spot diameter on the sample was about 500 $\mu m$) and averaged to minimize systematic errors.

\subsubsection*{Disentanglement of the MLD and structural dichroism \label{Data}}

In our experimental geometry, with $x$/$y$ corresponding to
horizontal/vertical light polarizations, the MLD, given by $\epsilon
_{xx}-\epsilon_{yy}$ in Eq. (1), should result in a difference between
reflectivities of the \textit{s-} and \textit{p-}polarized light. However, the
weak modulation of the intensity of the reflected light is also expected due
to effects related to the operation of the PEM, as well as due to
structural/strain anisotropy. Thus, the modulation of light intensity on the
photo-diode, $\Delta I=\frac{\Delta R}{R}$ (the reflectivity, $R$, of Mn$_{2}$Au at
852 nm is 0.52 +/- 0.01), can be expressed as:%

\begin{equation}
\Delta I_{i}(\phi)=D_{0}P(\phi)+D_{mag}^{i}P(\phi)+D_{str}^{i}P(\phi).
\label{Signal}%
\end{equation}

Here $D_{0}$ is the contribution to the signal related to the PEM operation
and slight misalignment of optical components (present also when measuring
uncoated gold mirror as a reference). $D_{mag}^{i}$ is the amplitude related
to the MLD of the sample, where $i=A\ $- $C$. Finally, $D_{str}^{i}$ is the
amplitude of the dichroism caused by structural effects, such as strain. Here,
we assume that in the first approximation the structural/strain dichroism is
decoupled from $D_{mag}^{i}$, \textit{i.e.} the strain does not affect the
N\'{e}el vector. Thus, we assume $D_{str}^{i}$ is independent on the N\'{e}el
vector direction, but depends on the sample orientation (see also Appendix B). Finally, $P(\phi)$ is
a periodic function describing the polarization state of light after passing
through the PEM - see Fig. 2(a). Since identical experimental geometry is used
for all samples, $D_{0}$ is the same in all measurements. Moreover, given the
samples $A$-$C$ were cut from the same film, $D_{str}^{i}$ should be identical
to all samples, when mounted along the same direction with respect to the
as-grown sample. Conversely, $D_{str}^{i}$ should change sign when the sample
is rotated by 90$^{o}$ around the C$_{4}$ axis, \textit{i.e.} $D_{str}%
^{i}=-D_{str}^{i(90)}$. When considering the MLD, $D_{mag}^{A}\approx0,$ given
the multidomain state on the length scale of the laser spot size of 500 $\mu
$m. Finally, taking into account orthogonal direction of the N\'{e}el vector
in the two magnetically aligned samples,it follows that $D_{mag}^{B}=-D_{mag}^{C}$. Under
these assumptions the structural and magnetic contribution can be disentangled
by the following referenced measurements:%

\begin{align}
D_{mag}P(\phi)/2 =\Delta I_{B}(\phi)-\Delta I_{A}(\phi)=\Delta I_{A}(\phi)-\Delta I_{C}(\phi)%
\label{Mag}\\
D_{str}P(\phi) =\Delta I_{A}(\phi)-\Delta I_{A}^{(90)}(\phi)=\Delta I_{B}(\phi)-\Delta
I_{C}^{(90)}(\phi) \label{Str}%
\end{align}

\subsubsection*{Near-infrared magnetic linear dichroism in Mn$_{2}$Au}

Fig. 3 presents the results of dichroism measurements, which are performed in
sequence, under identical conditions. The dashed black line, representing the
evolution of the polarization state - see Fig. 2(b) - is added to all panels as
a guide to the eye. Fig. 3(a) presents the raw data, $\Delta I_{i}(\phi)$, taken
on samples $A$, $B$ and $C$. The complicated shape of the raw signal is
largely a result of the optical elements of the setup, mainly the PEM. The PEM
acts as a $\lambda\cdot\delta$ wave plate, where $\delta$ is modulated between
$-1/2$ and $1/2$. This results in a sequence of polarization states
\textit{s-p-s-p-s} within one modulation period of the PEM (20 $\mu$s). Note
that between \textit{s-} and \textit{p-} polarizations, the light is
circularly/elliptically polarized. Due to imperfections and asymmetry between
the optical properties of the squeezed and elongated PEM crystal a modulation
of the signal at 50 kHz is also observed.

\begin{figure}[htp]
\centering \includegraphics[width=0.95\columnwidth]{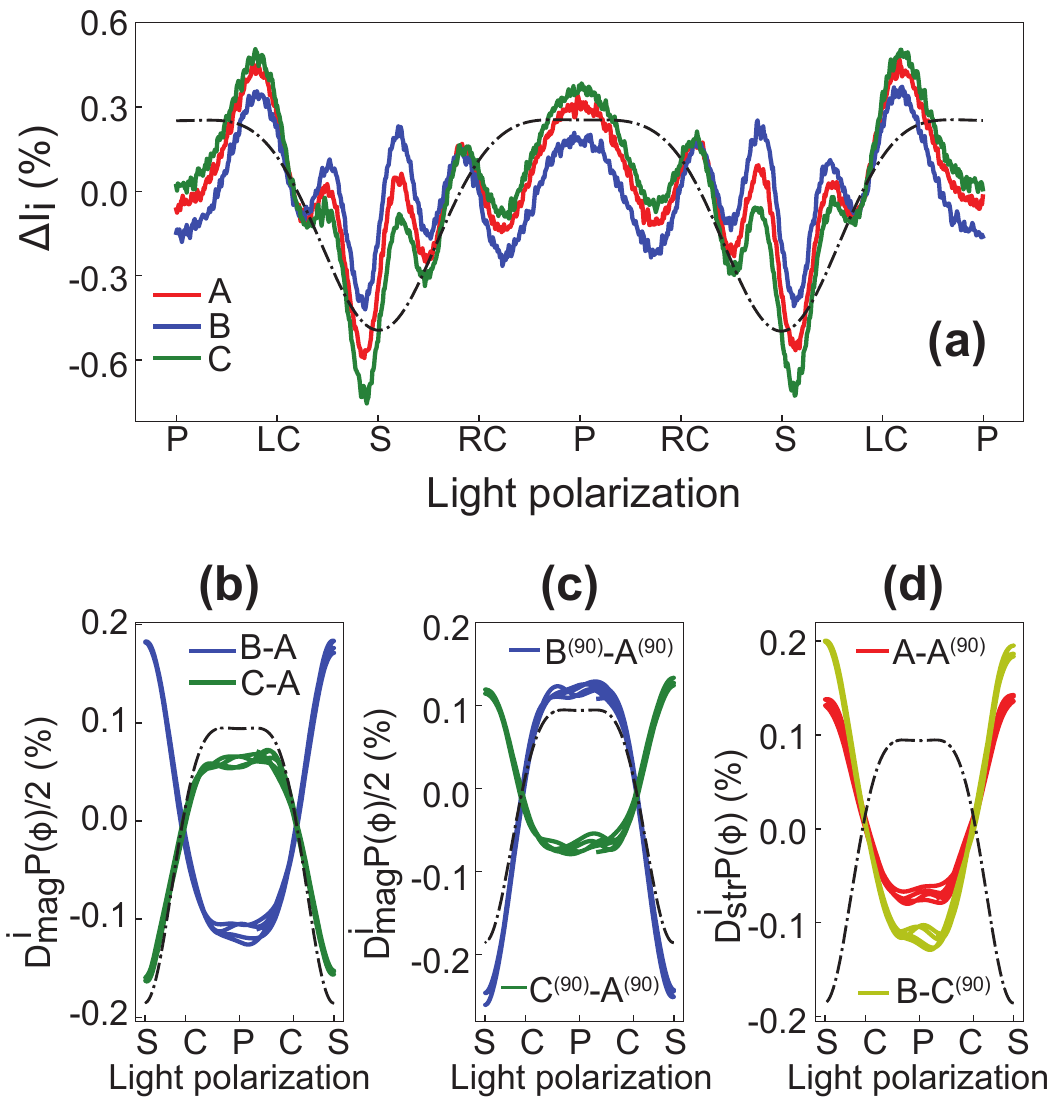}\caption{Results
of reflectivity anisotropy studies of Mn$_{2}$Au films with different orientations of the magnetic order. (a) The raw reflectivity modulation signals from all three samples, oriented in the same growth direction (raw data obtained on samples rotated by 90$^{o}$ are not shown). Here S, P, LC and RC stand for the \textit{s}-polarized, \textit{p}-polarized, and left and right circularly polarized light, respectively. The time window here is 20 $\mu$s. Panels (b)-(d) present the differential signals, disentangling the MLD and structural contributions to reflectivity variation according to Eqs. (\ref{Mag}) and (\ref{Str}). Here the traces have been folded into the 10 $\mu$s time window to emphasize the lack of circular dichroism, which should have a 50 kHz component. The modulation in (b) and (c) is a result of the MLD, while the modulation in (d) reflects the structural component to dichroism. Note that the modulation of reflectivity was recorded using a digital oscilloscope in the AC coupling mode. Thus, the zero baseline is automatically determined by the oscilloscope as the mean value of the periodic signal. Due to the asymmetry of the signal, the zero level does not correspond to the arithmetic average of the maximum and minimum. The asymmetry is a result of PEM operation, where the flat regions (here the light is p-polarized - see Figure 2(b)) correspond to the maximal retardation of the photo-elastic modulator.}%
\label{Dichroism}%
\end{figure}

To obtain the MLD component we first subtract the signal of the reference
sample (as-prepared, $A$) from signals obtained on the magnetically polarized
samples $B$ and $C$, following Eq.(\ref{Mag}). Moreover, as the MLD should give
rise to a periodic modulation in reflectivity at the second harmonic of the
PEM frequency, we fold the data recorded over two periods into the 10 $\mu$s
time window. The resulting variations are shown in Fig. 3(b). The differential
signals do not have a complicated shape of the raw data as in Fig. 3(a),
demonstrating that subtraction efficiently cancels out all system related
modulations. Indeed, the shape of the differential signal follows the shape of
the reference curve (dashed line). Furthermore, the absence of modulation of
the differential signal at the 50 kHz PEM frequency clearly confirms the
absence of circular dichroism. Fig. 3(c) presents the result obtained by the same approach for samples rotated by 90$^{o}$ around the C$_{4}$ axis. The fact that the differential signals
$\Delta I_{B}^{(90)}(\phi)-\Delta I_{A}^{(90)}(\phi)$ and $\Delta I_{C}^{(90)}(\phi)-\Delta I_{A}^{(90)}(\phi)$ in Fig. 3(c) are phase-shifted by $\pi$ underscores the MLD nature of the differential signals. 

The MLD-induced change in reflectivity is given by the differences between the maxima and minima of the traces in Fig. 3 (b-c), multiplied by a factor of two (the signals shown in Fig. 3(b) and (c) are relative to the unpolarized reference sample $A$). The reflectivity is found to be higher for the light polarized parallel to the staggered magnetization. The average value of the MLD induced reflectivity change from all data combined is $0.52 \pm 0.14 \%$. We note, however, that it follows from Fig. 3(b) and (c) that the MLD extracted from $\Delta I_{B}(\phi)-\Delta I_{A}(\phi)$ and $\Delta I_{B}^{(90)}(\phi)-\Delta I_{A}^{(90)}(\phi)$ ($0.63 \pm 0.03 \%$), is larger than the MLD extracted from $\Delta I_{C}(\phi)-\Delta I_{A}(\phi)$ and $\Delta I_{C}^{(90)}(\phi) - \Delta I_{A}^{(90)}(\phi)$ ($0.41 \pm 0.07 \%$). This may suggest that the underlying microstructural strain favors the orientation of the N\'{e}el vector parallel to the $c$-axis tilt (parallel to the $[010]_{s}$ axis of the substrate), resulting in a partial polarization of the as grown reference sample $A$ along the $[010]_{s}$ axis. Indeed, preliminary study using scanning electron microscopy with polarization analysis seem to support this observation \cite{SEMPA}. 

Fig. 3(d) presents results of the measurements of structural dichroism.
Following Eq.(\ref{Str}), for differences $\Delta I_{A}(\phi)-\Delta
I_{A}^{(90)}(\phi)$ and $\Delta I_{B}(\phi)-\Delta I_{C}^{(90)}(\phi)$ the signal
due to the MLD should cancel out and the remaining variation of reflectivity
is due to the structural/strain effects. The extracted $D_{str}=0.26 \pm 0.08 \%$, with the reflectivity being larger for light polarized perpendicular to the direction of the $c$-axis tilt, i.e. parallel to the $[211]_{s}$ axis of the r-cut sapphire  (see also Appendix B). Thus, the structural dichroism is about twice smaller than the MLD in Mn$_2$Au films on r-cut sapphire. Noteworthy, the value of $D_{str}$ extracted from $\Delta I_{A}(\phi)-\Delta I_{A}^{(90)}(\phi)$ is again noticeably lower that the value extracted from $\Delta I_{B}(\phi)-\Delta I_{C}^{(90)}(\phi)$. This is consistent with the above suggestion that the as-grown sample is partially polarized along the $[010]_{s}$ axis of the substrate, resulting in a lower value of the extracted structural dichroism when monitoring $\Delta I_{A}(\phi)-\Delta I_{A}^{(90)}(\phi)$.

\section{Conclusions}

We demonstrate a pronounced magnetic linear dichroism in the NIR range in a
collinear metallic antiferromagnet Mn$_{2}$Au at room temperature. The observed MLD of $\approx 0.6\%$ exceeds the value of AMR in Mn$_{2}$Au \cite{Bodnar_AMR}, recorded at liquid Helium temperatures, by about a factor of four. While no study on temperature dependence of the AMR in Mn$_{2}$Au has been performed thus far, one may expect that the increased electron-phonon and electron-magnon scattering give rise to a suppression of AMR at higher temperatures, as typically observed. Thus, the ratio between the MLD and AMR may in fact be even higher at room temperature.

Recent broadband terahertz study of several ferromagnets demonstrated AMR signals up to 30 THz \cite{Nadvornik}, with the effect slowly decreasing with increasing frequency. The
large difference between the observed MLD in the near-infrared and the AMR in
Mn$_{2}$Au \cite{Bodnar_AMR} implies the MLD in the near-infrared to be a
result of the polarization dependent interband absorption. Indeed, the recent
angular-resolved photoemission spectroscopy study does show that the N\'{e}el
vector orientation leads to a pronounced valence band asymmetry up to binding
energies of several eV \cite{harpes_mn2au}. Thus, further systematic studies of MLD
as a function of photon energy may reveal spectral ranges with even higher MLD.

In view of the recent studies, demonstrating a large magneto-optical contrast in ultrathin NiO \cite{Xu19,Schreiber,Meer} and CoO \cite{Xu20} films, it is tempting to estimate the polarization rotation angle, which corresponds to the observed MLD-induced reflectivity variation of $\approx 0.6\%$. Assuming the observed reflectivity change is dominated by the variation of the (interband) absorption coefficient, we estimate the polarization rotation angle for the light polarized at 45$\circ$ away from the N\'{e}el vector to be $\approx 340$ mdeg, about a factor of 2 larger than the reported value in CoO thin films \cite{Xu20} and more than a factor of 5 larger than the reported value in NiO thin films \cite{Xu19}.

As a simple non-perturbative table-top experiment, the presented approach can be extended to
imaging mode, enabling detection of N\'{e}el vector orientation with
micrometer spatial resolution. Moreover, in view of potential applications of
Mn$_{2}$Au, \textit{e.g.} as a spintronic memory device, utilizing sequences
of femtosecond optical pulses can provide read-out speeds matching the
expected ultrafast switching times in antiferromagnetic memory devices.

\begin{acknowledgments}
This work was funded by the Deutsche Forschungsgemeinschaft (DFG, German
Research Foundation) Grant No. TRR 173 268565370 (project A05). This work
received support from Horizon 2020 Framework Program of the European
Commission under grant agreement No. 863155 (S-NEBULA). V.G. and M.F.
acknowledge the financial support from the Graduate School of Excellence
"Materials Science in Mainz" (DFG GSC 266 49741853). We acknowledge the Paul Scherrer Institut, Villigen, Switzerland for the beamtime allocation under proposal 20200977 at the SIM beamline of the SLS. The authors thank the SIM beamline staff for the technical support. We acknowledge
valuable discussions with H.-J. Elmers, H. Gomonay, P. Grigorev and L. \v{S}mejkal.
\end{acknowledgments}

\section*{Appendix A: Experimental setup for optical dichroism}

Our experimental setup is sketched in Fig. 2(a). For the non-cubic material, the
experimental reflectivity-anisotropy can be a result of a finite angle of
incidence. For the same reason, \textit{i.e.} the difference in reflectivities
of $p$- and $s$-polarized light, any dielectric mirrors need to be avoided in
the beam path. Therefore, it is not possible to use normal incidence and to
separate the reflected light using a semi-transparent dielectric mirror. To
minimize such effects, the angle of incidence of $\alpha/2\approx0.34^{o}$ is
used. Moreover, we use minimal number of the optical elements, with the only
reflecting surface being the surface of the sample, as shown in Fig. 2(a).

We use the 852 nm laser diode and a Photo-Elastic modulator (PEM) from Hinds
Instruments, which operates at the frequency of 50 kHz. The PEM is set to
$\lambda/2$ retardation mode for 852 nm, resulting in a modulation of light
polarization between $s$-polarized (vertical) and $p$-polarized (horizontal)
at 100 kHz, the second harmonic of the PEM - see Fig. 2(b).

The laser intensity was 5 mW while the diameter of the laser spot on the sample was 500 $\mu$m. The laser heating in thin films is governed by the absorbed light intensity and the thermal properties of the substrate (sapphire in our case). The temperature increase of the illuminated region can be estimated using a simple steady-state heat diffusion model \cite{heating,mihailovic}. Using the reflectivity of 0.52, thermal conductivity of sapphire at room temperature of 35 W/mK \cite{kappa} we estimated, using Eq.(3) of Ref.\cite{mihailovic}, the steady state heating to be $<0.1$ K. Thus, the laser heating can be neglected.

\section*{Appendix B: The possible origins of structural dichroism}

While our data clearly demonstrate the presence of a substantial MLD in
Mn$_{2}$Au, there are two possible origins giving rise to the structural
component of the measured reflectivity anisotropy - see Fig. 3(d). To address
these, we first need to consider the specifics of the epitaxial film growth on
r-cut sapphire using the Ta (001) buffer layer \cite{Jourdan_growth,FilmsAFM}.

\begin{figure}[htp]
\centering \includegraphics[width=0.95\columnwidth]{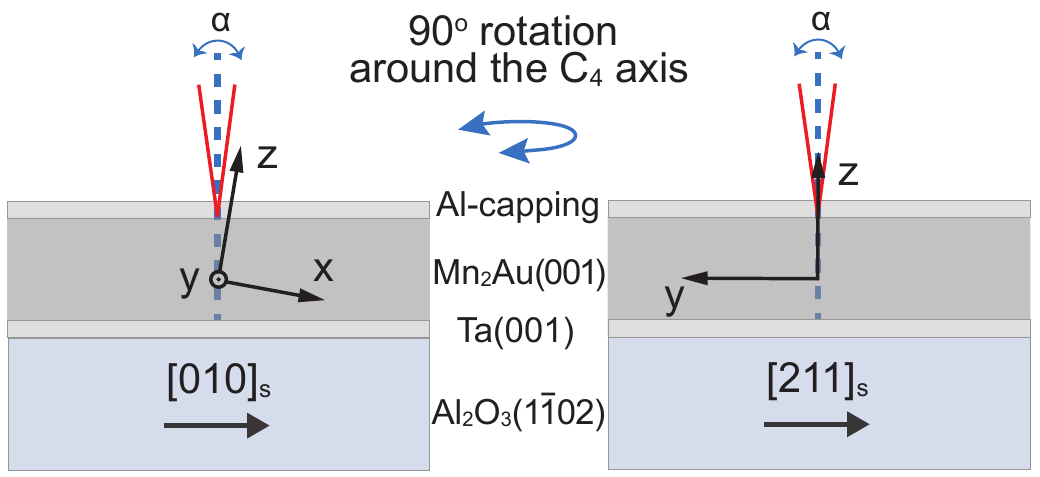}\caption{Schematic
of the sample layout in dichroism measurements for the two different
orientations of the sample (left and right) with respect to the laser beam
directions (red lines). Here $x$ and $y$ correspond to [110] and
[1$\overline{1}$0] directions of Mn$_{2}$Au, while $z$ is parallel to the
$c$-axis, [001]. The tilt of the $c$-axis is highly exaggerated.}%
\end{figure}

As elaborated elsewhere \cite{Bodnar_PRB, Bodnar_thesis}, the $c$-axis of the
Mn$_{2}$Au films grown on (1$\overline{1}$02) Al$_{2}$O$_{3}$ is tilted with
respect to the surface normal by about 2-3$^{o}$ (the tilt shows small
variation from sample to sample). It turns out the the $c$-axis of the film
tilts towards the $[010]_{s}$ axis of the substrate.

Given the tilted growth, the observed structural dichroism could be a result
of the experimental geometry - the effect is illustrated in Fig. 4. Namely,
due to the tilt of the film's c-axis with respect to the surface normal, the
rotation of sample by $90^{\circ}$ (or equivalently by rotating the light
polarization) changes the projection of the $\vec{k}$ vector of light onto the
\textit{c}-axis. The difference between $\epsilon_{\bot}$ and $\epsilon_{||}$
can thus give rise to reflectivity modulation. Unfortunately, due to the lack
of data on (anisotropic) optical properties of Mn$_{2}$Au, this contribution
cannot be estimated.

On the other hand, the observed dichroism can also be a result of the
underlying strain as a result of the anisotropic thermal expansion of sapphire
\cite{Saphire1,Saphire2}; such an effect can not be neglected on the
low-symmetry cut of the Al$_{2}$O$_{3}$. Indeed, investigations of sample
morphology using atomic force microscopy suggest breaking of the four-fold
symmetry \cite{FilmsAFM}. Considering the underlying strain to be responsible
for the non-magnetic components of dichroism, its amplitude could roughly be
estimated from $D_{str}$. Assuming a linear relation between $D_{str}$ and the
relative difference in the $[110]$/$[1\overline{1}0]$ lattice spacings the
latter should be of the order of 0.1 \%. Using the XRD measurements we are,
however, unable to resolve such a weak asymmetry between the [110] and
[1$\overline{1}$0] lattice spacings parameters, because of the rather large
mosaicity of $\approx0.5^{o}$ \cite{Jourdan_growth,FilmsAFM}.

\end{document}